# NFC based inventory control system for secure and efficient communication

Razi Iqbal[1], Awais Ahmad[2], Asfandyar Gillani[3]

*Department of Computer Science and IT, Lahore Leads University1, 3*
*Al-Khawarizmi Institute of Computer Science, University of Engineering and Technology2,*
*razi.iqbal@leads.edu.pk1, awais.ahmad@kics.edu.pk2, hod.cs@leads.edu.pk3*

## ABSTRACT

This paper brings up this idea of using Near Field Communication (NFC) for inventory control system instead of using traditional barcodes. NFC because of its high security, ease of use and efficiency can be very suitable for systems like inventory control. In traditional inventory control systems, each product has a barcode pasted on it, which is vulnerable to attacks as barcodes are open and have no security. Furthermore, barcodes are prone to damages and can be unreliable when pasted on different types of products e.g. hot and frozen products, circular shaped products and irregular shaped products like clothes etc. NFC on the other hand is very efficient, secure and reliable when it comes to short-range wireless communication. In this paper we will present our prototype for the inventory control system of an electronic store in which each product has a passive NFC tag pasted to it. When a customer buys a product the receipt of the product is generated using NFC between the NFC passive tag on the product and NFC enabled device (e.g. smart phone or reader) at the cash counter.

## 1. INTRODUCTION

The paper discusses the use of NFC and Barcodes for improving the user interaction and service discovery in mobile devices. [1] [2]. An Inventory control system is about controlling, managing and processing the products in the inventory. It is an automated system, which not only reduces human effort, cost and mistakes but also increases the efficiency and speed at which items can be processed in an inventory. Majority of inventory control systems throughout the world use barcode or RFID (Radio Frequency Identification) for identification of items in an inventory. In the case of a barcode based inventory control system, a barcode is pasted on each product. When a customer selects the product and goes to the cash counter to pay for the product, the cashier scans the barcode using a barcode reader, which displays the information of the product on the screen, and that product is marked as sold. Figure 1 gives a general overview of how barcode and RFID based inventory control systems work.

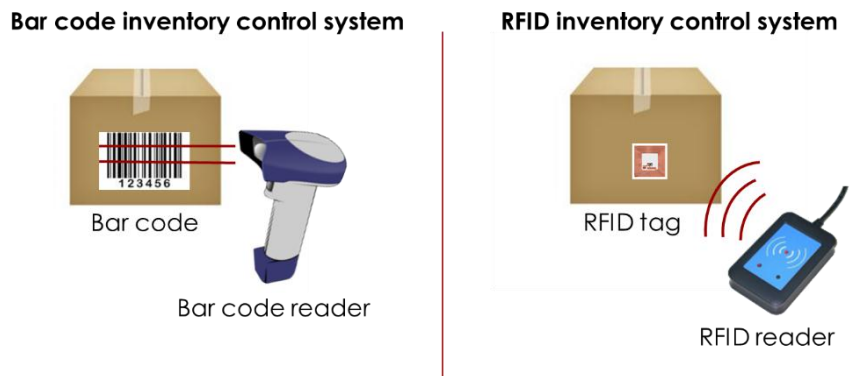

FIGURE 1. Traditional Inventory Control Systems

Majority of the inventory systems throughout the world are based on barcode because of its simplicity to use. RFID on the other hand are a bit complex, needs extra effort, error occurrence because of collision of multiple tags when read at the same time and one way communication between the tag and the reader.

This paper presents the idea of using Near Field Communication (NFC) for inventory control system instead of barcode based inventory control systems because of reliable, efficient and secure communication of NFC. The paper will explain in detail our prototype system of NFC based inventory control system for an electronic store where each product will have a NFC tag pasted on it and all the inventory management is done using NFC.

## 2. RESEARCH ORIGINALITY

With the increase in the use of technology in every form of life, the trend of managing and controlling products in an inventory has also changed. Items are no longer managed manually on registers and instead the process is automated which has increased the efficiency and processing speed of the whole process. Near Field Communication is the next big thing in short range wireless technologies. With its secure short range and highly efficient data transfer techniques, NFC has the potential to be used in different domains like education, entertainment, transportation and many more. Till date, many NFC enabled systems have been developed which are far more efficient and reliable than old traditional systems. Our research on NFC has enabled us to use NFC in inventory control system and develop such a system, which is efficient, reliable, secure and cost effective than already available systems. NFC has never been used in inventory control system although it is fully eligible for it, so our system is one of a kind till date. Various aspects of NFC have been explored by conducting different experiments and its comparison with other sister technologies show that NFC based inventory control system can be very reliable.

## 3. SYSTEM MODEL

Figure 2 shows the system model for NFC based inventory control system for an electronic store. It uses a system of Internet of things and uses NFC/barcodes for interaction with the system [3]. As shown in the figure, each product in the store contains an NFC passive tag. A NFC passive tag is like a paper sticker, which is designed in such a way that it contains an NFC chip and the antenna. NFC passive tags are called passive tags because they do not have their own power to operate;

instead they are powered by the magnetic field generated by the active tag (NFC reader). An active tag is called active because it has its own power to operate; that can be battery or USB powered depending upon the type of the reader.

When a customer selects a product to purchase, for example a laptop, the cashier at the cash counter will touch the passive NFC tag pasted on the laptop with an active NFC reader at the cash counter; the NFC reader reads the information and appropriate action can be taken. One of the major advantages of NFC over bar code is the storage capacity of NFC tags. NFC tags can hold data up to 4KB, which is sufficient to hold the details for a product, like, product id, name, price, expiry date, manufacturing date and delivery date etc. NFC tags can actually hold all this data within itself instead of storing the data in the databases and then retrieving it at the time of payment at the cash counter. This will significantly reduce the processing time at the cash counter.

However, to make the comparison of NFC based inventory control systems with already available bar code inventory control systems; we will store the data in the database and then retrieve it from the database when a reader reads the passive NFC tag. Furthermore, to reduce the cost of the system we will use less storage NFC passive tags having storage capacity of around 128 bytes, which is still far more than barcode capacity. NFC and barcode characteristics are compared in detail in upcoming sections.

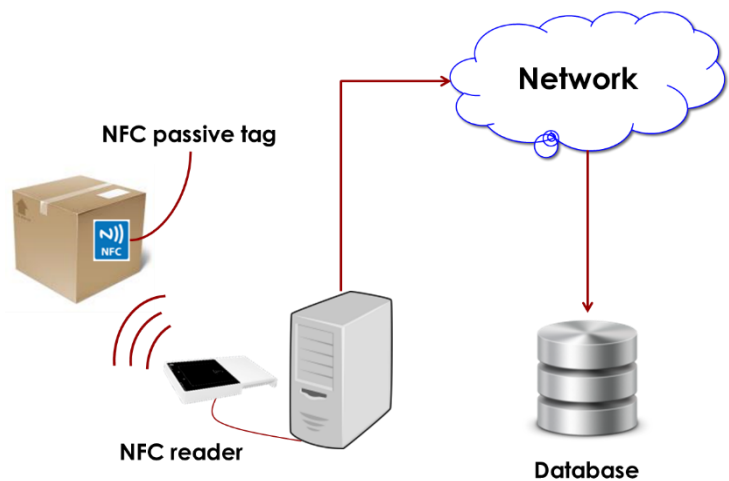

FIGURE 2. System Model for NFC based Inventory Control System

## 4.  SYSTEM FLOW

Figure 3 shows the system flow for proposed NFC based inventory control system. As illustrated in the figure above, each product has an NFC tag attached to it. The tag contains information about the product e.g. product id and name etc. At the cash counter, an NFC reader device is attached with a computer, which is connected to the network. As customer brings the product to the cash counter, the cashier at the cash counter will swipe the product containing NFC tag against the NFC reader. The information about the product will be read using NFC and is checked against the network database. If the product is available, proper receipt will be made and item will be marked as sold. If however, the product is not available, an error will be generated.

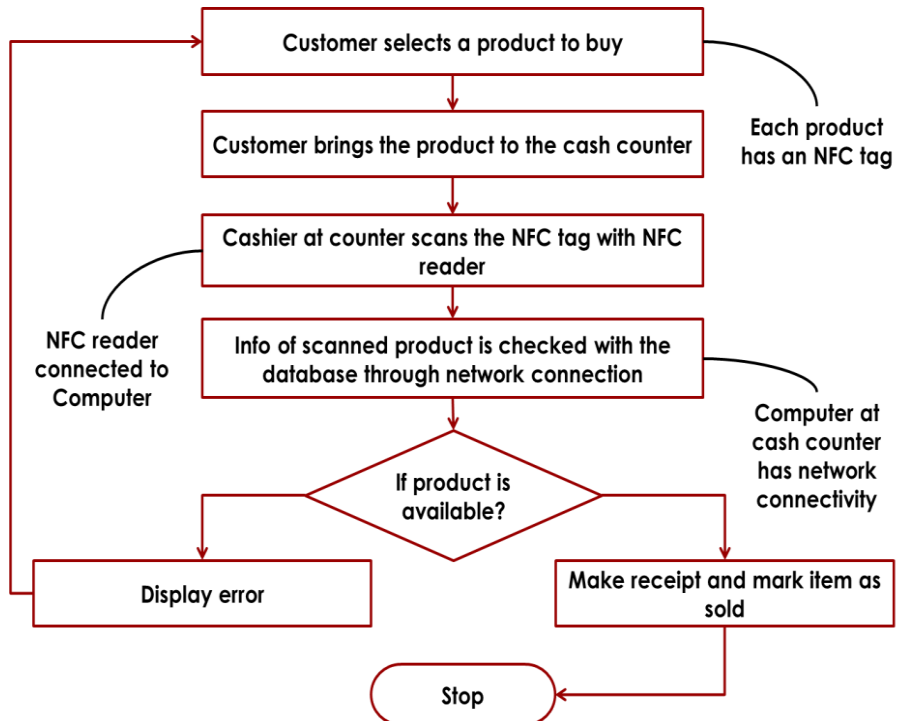

FIGURE 3. System flow for NFC based Inventory Control System

## 5. WHY USE NEAR FIELD COMMUNICATION?

The worldwide penetration of mobile phones at present makes mobile phones excellent devices for delivering new services to users without requiring learning effort [3]. An NFC-enabled mobile phone will allow a user to demand and obtain services by touching its different elements in a given smart environment. In this paper, we present a proposal in which we analyze the scope of touch interaction and develop a perceived touch interaction through tagging context (PICTAC) model. [4]

It is an emerging technology for communication within a few centimeters. Because of its low range it has high security. It's a simple short range wireless technology with 2 way communication between electronic devices. NFC is based on open and maintained standards like ISO (International Standard Organization), ECMA (European association for standardizing information and communication systems) and ETSI (European Telecommunications Standards Institute). NFC is maintained by NFC-Forum, which was formed, by Sony, Phillips and Nokia back in 2004. NFC-Forum has more than 200 members including developers, institutions and manufacturers.

The major benefit of using NFC is its simplicity. A user can use it simply by touching a tag and the action is performed, without worrying about the complex process going on in the background, and this is what commercial technologies are all about; bringing simple solutions for the masses.

NFC is low power, around 15ma and a low data transfer rate of around 424kbps, which is sufficient for lightweight information transfer like secure financial transactions but too slow for media streaming like video playback etc. As NFC is not built for streaming heavy amount of data and is mainly for lightweight data transfer, so low data rate of NFC is justified. Similarly NFC is ideal for situations where instant connection is required, for example e-ticketing at Metro Stations.

TABLE 1.
Near Field Communication comparison with other technologies

| Item | NFC | Bluetooth | IrDA |
| --- | --- | --- | --- |
| Data Rate | 424kbps | 1Mbps | > 10Mbps |
| Range | 4-10cm | 5-10m | < 1m |
| Ease of Use | Very easy | Difficult | Difficult |
| Availability | Easy | Easy | Easy |
| Instant Pairing | Yes | No | No |
| Encryption | Yes | Yes | Yes |
| Two way communication | Yes | Yes | Yes |

## 6. NFC VS. BAR CODE INVENTORY CONTROL SYSTEMS

This section discusses the use of NFC and barcodes in supply chain and inventory management [5]. As mentioned in the previous section, NFC tags are very efficient and secure as compared to tags of other technologies like bar code [6] [7]. Figure 4 shows the working model of bar code and NFC based inventory control systems.

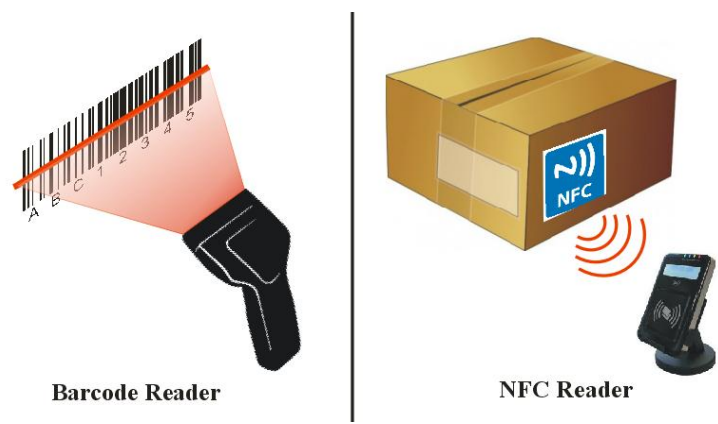

**Barcode Reader**                    **NFC Reader**

FIGURE 4. Barcode vs. NFC inventory control systems

This paper presents the comparison of electronic store inventory control systems based on barcode and NFC. We have developed two different inventory control systems; one based on barcode and other based on NFC. Different experiments have been performed to determine the efficiency, reliability, readability, affordability and security of both the systems. Below are details of each attribute with comparison of both the technologies: -

### 6.1 Readability

Barcode tags are very efficient if they are read in such a way that the whole barcode is read at a specific angle. The readability of the barcode reduces significantly if the barcode reader is not aligned properly with the barcode itself. Figure 5 below shows how a barcode can and cannot be read properly by the barcode reader.

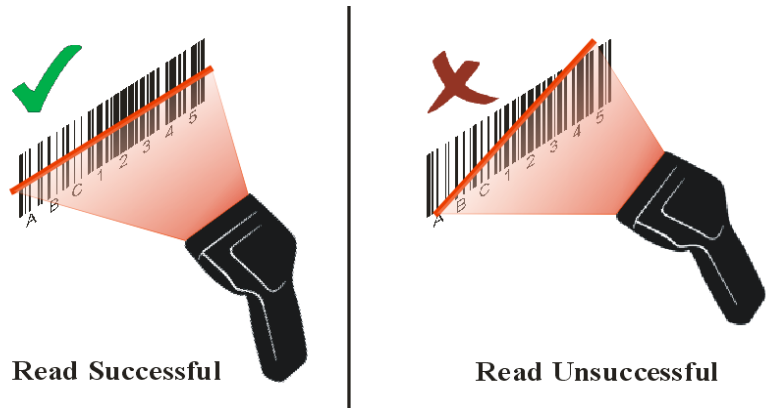

FIGURE 5. Barcode readability at different angles

As shown in Figure 5 above, the barcode reader should face the barcode in such a way that the whole barcode must be exposed to the barcode reader. Even if a single bar of barcode is not properly exposed to the barcode reader, the whole barcode will not be read. This fact reduces the efficiency of the barcode significantly.

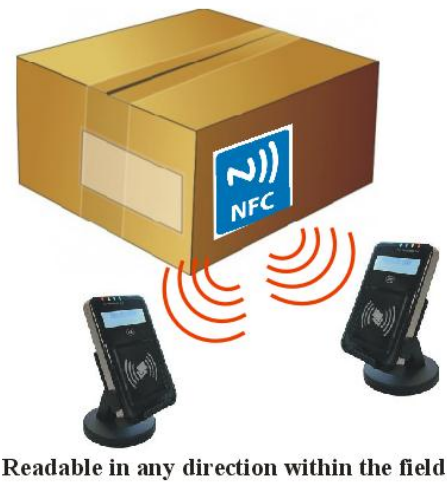

Readable in any direction within the field

FIGURE 6. NFC readability at different angles

However, as compared to barcode, NFC readers can read NFC tags at any angle. NFC readers do not need to be in proper alignment with NFC tags to retrieve the information, which significantly increases the efficiency of the inventory control systems. Figure 6 above shows that communication angle between NFC tags and readers do not matter as long as tags are in reading range of the readers, which is around 10cm.

Different experiments have been conducted to determine the readability of barcode and NFC. Table 2 below shows the experimental data for the readability for both barcode and NFC.

TABLE 2.
NFC and barcode readability experiments

|          | Angle (Degree) | Readability  |
|----------|----------------|--------------|
| Barcode  | 0              | Readable     |
|          | 1              | Readable     |
|          | 2              | Readable     |
|          | 3              | Readable     |
|          | 4              | Readable     |
|          | 5              | Readable     |
|          | 6              | Readable     |
|          | 7              | Readable     |
|          | 8              | Readable     |
|          | 9 - 172        | Not Readable |
| NFC      | 0-360          | Readable     |

The data in the table above shows that barcode is not readable at all angles. Our experiments show that barcode is not readable if reader is tilted beyond 8-degree angle. However, NFC tags are readable no matter how much the reader is tilted, as long as the tag is in the field of the NFC reader [8].

## 6.2 Storage Capacity and Size

Each tag has some storage capacity in both barcode and NFC. There are limitation of storage capacity in tags in both barcode and NFC. In barcode, different standards are available, containing 8, 9, 13 and 14 characters. Parity check is applied to these standards for accuracy check. However with the increase in the number of characters, the size of the barcode tag will increase as shown in figure below:-

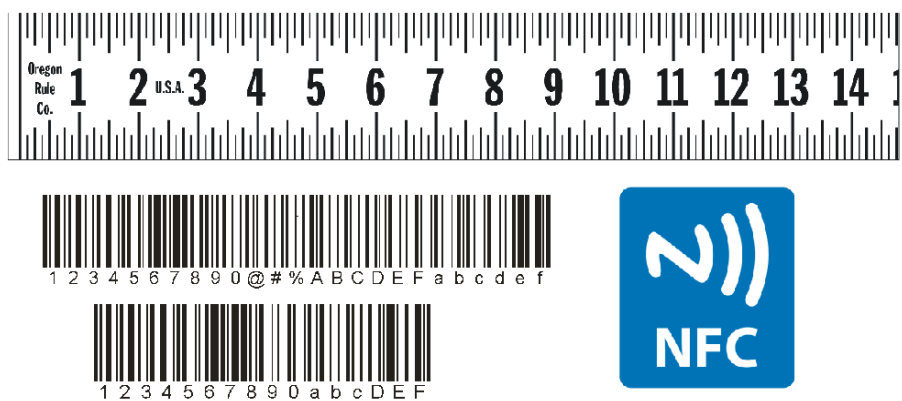

FIGURE 7. Barcode and NFC storage capacity

Increase in the size of the barcode tag reduces readability of the tag. A significant decrease in the readability occurs with the increase in the size of the barcode tag, which reduces the efficiency. Figure 8 shows the effect of size of the tag on readability.

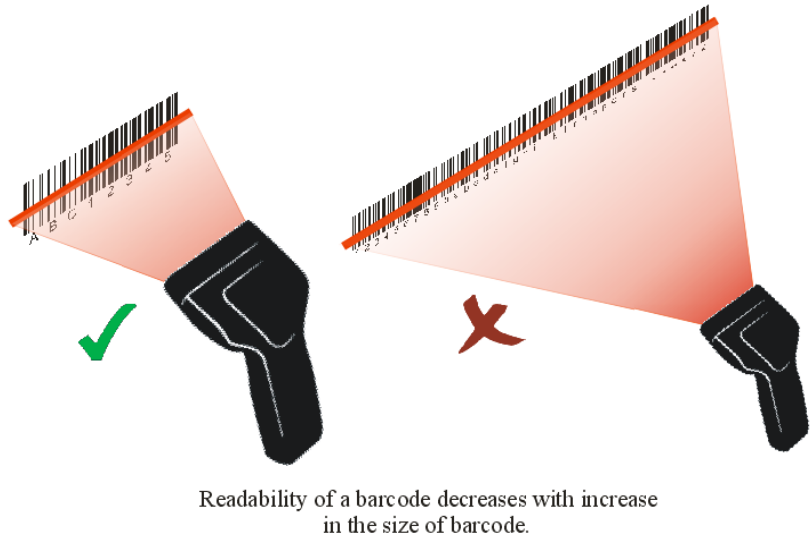

Readability of a barcode decreases with increase
in the size of barcode.

FIGURE 8. Decrease in readability with increase in size of barcode tag

In contrary to barcode, NFC tag size is not directly proportional to storage capacity. The size of the NFC tag does not increase with the increase in the number of characters [9]. In NFC, different categories of tags are available, each having its own storage capacity. For the purpose of this paper, we have selected 'NFC tag type 2' which has a maximum capacity of 128 bytes (128 characters). The size of NFC tag will remain same whether 2 bytes are used or 128 bytes are used. NFC tag is shown in Figure 7 above. Since the size remains the same, readability of the tag will remain constant and will not be affected by the increase in the storage capacity [10].

Table 3 below shows the experiments conducted for the comparison of NFC and barcode readability with respect to the size of the tag. The results show that with the increase in the storage capacity of barcode, the physical size of the tag increases and it gets difficult to read. However in the case of NFC, the size of the tag remains the same even if the storage capacity of the tag is increased which makes NFC tags more readable.

TABLE 3.
NFC and barcode readability experiments with respect to size

|         | No. of characters | Size      | Readability    |
| ------- | ----------------- | --------- | -------------- |
| Barcode | 8                 | 33mm      | Easy           |
|         | 12                | 35mm      | Easy           |
|         | 20                | 66mm      | Difficult      |
|         | 30                | 94mm      | Very Difficult |
| NFC     | 8                 | 30-40mm   | Easy           |
|         | 20                | 30-40mm   | Easy           |
|         | 128               | 30-40mm   | Easy           |

## 6.3  Risk of damages

Barcodes are susceptible to damages. Since the barcodes are printed, so they can be damaged intentionally or unintentionally. A little disturbance in the barcode increases the chances of barcode not being read.

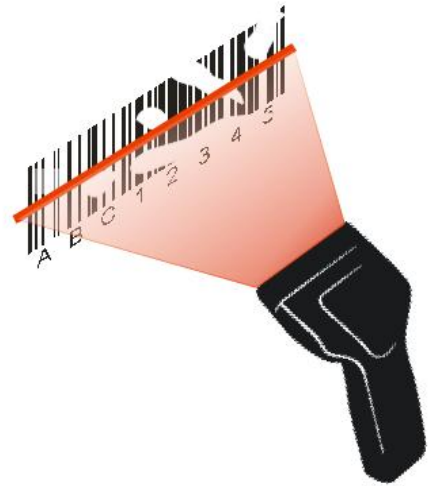

**A damaged/scratched barcode is not readable**

FIGURE 9. A damaged barcode has less chances of being read

As shown in Figure 9 above, if barcode is damaged or scratched, it won't be read by the reader, which makes barcode inefficient.

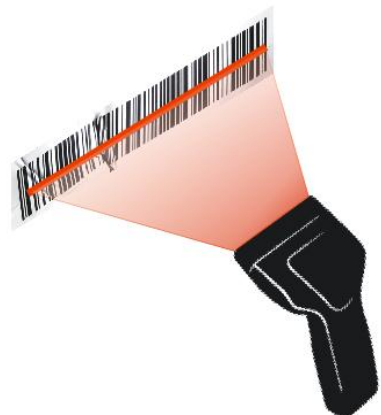

**Barcode Scanner is trying to read a wrinkled Barcode**

FIGURE 10. A wrinkled barcode has less chances of being read

Furthermore, as shown in Figure 10, since the barcodes are printed so there are chances that barcodes are wrinkled when pasted on rough surfaces. This wrinkled barcode is very difficult to read by the reader.

However, in NFC, since the tags are not printed, so there are not chances of data being erased. As long as the tag is exposed to the active antenna of reader, the data can be read which makes NFC a resilient choice as compared to barcode.

## 6.4 Security

Security is a major concern in today's environment. Data flow between the wireless modules can be compromised if not taken care of properly. In a barcode, security is not very strong as compared to NFC. Majority of barcodes are based on code39 algorithm which if cracker gets to know, whole system can be compromised [11].

NFC in combination with inexpensive passive tags is used to prevent attacks in a decentralized approach. The paper motivates the use of NFC for such a system and gives a short list of possible attacks. The underlying protocol to prevent attacks is shortly described and the requirements for all major components of such a system are defined. [12]. Furthermore, due to secure communication, NFC can be used in mobile payments as well and is considered as one of the better approaches available for mobile payment. NFC is being used in different countries for mobile payments because of its strong and secure architecture [13]. This ability of NFC makes it a good choice to be used in inventory control system where data is to be kept private.

## 6.5 Reusability

Barcodes are inexpensive and can be printed as per requirement. However once printed, barcodes are not rewritable. For example if the price of the product is changed, a new barcode should be printed to be used for the product and hence the old barcode is wasted. NFC tags, although little expensive than barcodes but are rewritable and can be used again and again with new information. This reusability of NFC tags can be very important where product information is updated frequently.

## 7. CONCLUSION

Our research showed that use of NFC in inventory control system could highly improve the efficiency in inventory control systems. The systems based on barcodes because of the limitations can be less efficient as compared to the systems based on NFC. Different experiments have shown that NFC has greater accuracy and readability as compared to barcodes since barcodes need to be read at a proper angle. Furthermore, increase in the storage capacity of NFC does not affect the size of the tag, which is one of the weaknesses of the barcode. This increase in the size of the barcode significantly reduces the readability of barcodes tags. However in the case of NFC, even if the storage capacity of the tag is increased, the size of the tag remains unchanged. Security is another factor, which makes NFC a better choice in inventory control systems than the barcode based inventory control systems, since NFC has multiple layers of security, which keeps the data secure and private if needed to be. These all attributes of NFC makes it's a good choice to be used in inventory control systems.